\documentclass[12pt]{article}
\usepackage{amssymb,amsmath,amsthm,latexsym}

\newtheorem{thm}{Theorem}[section]

\newtheorem{lem}[thm]{Lemma}

\newcommand{\pf}{{\bf Proof. \ }}

\begin{document}

\title{Linear complementary pair of group codes over finite principal ideal rings}
\author{Hualu Liu\\
 School of Science,\\
 Hubei University of Technology,\\
 Wuhan, Hubei 430068, China \\
{Email: \tt hwlulu@aliyun.com} \\
 Xiusheng Liu\footnote{Corresponding author.}\\
 College of arts  and science, \\
 Hubei Normal university  \\
 Huangshi, Hubei 435003, China, \\
{Email: \tt lxs6682@163.com} \\}
\maketitle


\begin{abstract}
A pair $(C, D)$ of group codes over group algebra $R[G]$  is called a linear complementary pair (LCP) if $C \oplus D =R[G]$, where $R$ is a finite principal ideal ring, and $G$ is a finite group.  We provide a necessary and sufficient condition for a pair $(C, D)$ of group codes over group algebra $R[G]$ to be LCP. Then we prove that if $C$ and $D$ are both  group codes over $R[G]$, then
$C$ and $D^{\perp}$ are permutation equivalent.
\end{abstract}


\bf Key Words  \rm  Finite principal ideal rings  $\cdot$ LCP of codes $\cdot$ Code equivalence

\bf Mathematics Subject Classification (2010) \rm 94B15 $\cdot$ 94B60 $\cdot$ 11T71

\section{Introduction}
Linear complementary pairs (abbreviated to LCP) of codes over finite fields whose are a class of special properties have been of interest and extensively studied due to their rich algebraic structure and wide applications in cryptography. LCP of codes was introduced by Carlet  et al. in \cite{Carlet} and \cite{Carlet1}, showed that these pairs of codes can  help improve the security of the information processed by sensitive devices, especially against so-called side-channel attacks (SCA)  and fault injection attacks (FIA). The most generic and efficient known protection against SCA is achieved with masking: every sensitive data (that is, every data processed by the algorithm from which a part of the secret key can be deduced) is bitwise added with a uniformly distributed random vector of the same length or several ones, called globally a mask. If the sensitive data and the mask belong respectively to two supplementary subspaces $C$ and $D$ of a larger vector space, it is possible to deduce the sensitive data from the resulting masked data. And it is shown that the level of resistance against both SCA and FIA depends on $d_{LCP}(C,D)=\mathrm{min}\{d(C),d(D^{\perp})\}$ which is called the security parameter, where $d(C)$ is the minimum distance of the code $C$ and $d(D^{\perp})$ is the dual distance of the code $D$. This method is called Direct Sum Masking (DSM), and the pair $(C, D)$ is called a complementary pair of codes. Note that the linear complementary dual (LCD) codes amount to the special case when $D = C^{\perp}$, in which case the security parameter is simply the minimum distance of $C$. We refer to \cite{Ngo} for further information on complementary pairs of codes over finite fields and their uses.

Let $\mathbb{F}_{q}$ be the finite field with $q=p^{m}$, where $p$ is a prime number and $m\geq1$ is an integer.  Carlet et al. \cite{Carlet1}
showed  that  if $(C, D)$ is LCP, where $C$ and $D$ are both cyclic or $2D$ cyclic codes of length $n$ over $\mathbb{F}_{q}$ and $\mathrm{gcd}(n,q)=1$, then  $C$ and  $D^{\perp}$ are permutation equivalent. In \cite{Cem}, they showed that the same result holds if $C$ and $D$ are $mD$ cyclic codes for $m \in \mathbb{N}$. If $G$ is any finite group, a right ideal of $\mathbb{F}_q [G]$ is called a group code. In \cite{Bor}, Borello et al. obtained the most general statement for any finite group (also without a restriction on the order of the group) by showing that if $(C, D)$ is LCP of group codes ($2$-sided ideals) in $\mathbb{F}_q [G]$, then $C$ and $D^{\perp}$ are permutation equivalent. Just recently, in \cite{Cem1},G$\ddot{u}$neri  et al. had extended this result to  finite chain rings. Namely, they had proven that for an LCP of group codes $(C, D)$ in $R[G]$, where $R$ is a finite chain ring and $G$ is any finite group, $C$ and $D^{\perp}$ are permutation equivalent. Note that in particular this implies $d(C) = d(D^{\perp})$. Hence, there is an LCP of $2$-sided group codes over finite chain rings which has as good as the security parameter the $2$-sided group code with the best minimum distance.

The purpose of this paper is to examine LCP of codes over finite principal ideal rings. In section 2, we recall the necessary background materials on finite  principal ideal rings, and give a decomposition of group algebra $R[G]$. In section 3, we first give a characterization of LCP of group codes over group algebra $R[G]$.  Then shown that if $C$ and $D$ are two group codes over $R[G]$, then $C$ and $D^{\perp}$ are permutation equivalent. Hence the security parameter for an LCP of  group codes $(C, D)$ is simply $d(C)$.

\section{Preliminaries}
In this section, we first recall definitions and properties of finite chain rings $\widetilde{R}$ and finite principal ideal rings $R$, necessary for the development of this work.  For more details, we refer to  \cite{Doug,Doug1,Liu,Nor}. Then we give a decomposition of group algebra $R[G]$ for finite group $G$.

A commutative ring  is called a {\it chain ring} if the lattice of all its ideals is a chain.
It is well known that if $\widetilde{R}$ is a finite chain ring, then $\widetilde{R}$ is a principal ideal ring and
has a unique maximal ideal $\mathfrak{m}=\langle\gamma\rangle=\widetilde{R}\gamma=\{r\gamma\,|\,r\in \widetilde{R}\}$. Its chain of ideals is

\begin{equation*}
\widetilde{R}=\langle\gamma^{0}\rangle\supset\langle\gamma^{1}\rangle\supset\cdots\langle\gamma^{e-1}\rangle\supset\langle\gamma^{e}\rangle=\{0\}.
\end{equation*}
The integer $e$ is called  the {\it nilpotency index} of $\mathfrak{m}$.
Note that the quotient $\widetilde{R}/\mathfrak{m}$
is a finite field $\mathbb{F}_q$,
where $q$ is a power of a prime $p$.

Throughout this paper, let $R$ denote a principal ideal ring, $\mathfrak{m}_1, \ldots ,\mathfrak{ m}_s$ denote the maximal ideals of $R$, $e_1, \ldots , e_k$ denote their indices of stability. Then the ideals $\mathfrak{m}_1^{e_1}, \ldots ,\mathfrak{ m}_s^{e_s}$ are relatively prime in pairs and $\prod_{i=1}^s\mathfrak{m}_i^{e_i}=\bigcap_{i=1}^s\mathfrak{m}_i^{e_i}=\{0\}$. By the ring version of the Chinese Remainder Theorem, the canonical ring homomorphism $\Psi: R \rightarrow \bigcap_{i=1}^s R/\mathfrak{m}_i^{e_i}$, defined by $r \rightarrow  (r+\mathfrak{m}_1^{e_1}, \ldots ,r+\mathfrak{ m}_s^{e_s})$, is an isomorphism. Denote the chain rings $R/\mathfrak{m}_i^{e_i}$ by $R_i$  for $1\leq i\leq s$. The maximal ideal $\mathfrak{m}_i/\mathfrak{m}_i^{e_i}$ of $R_i$ has nilpotency index $e_i$. Then
$$R= R_1\times R_2\times\cdots\times R_s.$$

Let $G=\{g_1,\ldots,g_n\}$ be a group and denote by $R[G]$ (or $R_j[G]$) the group ring of $G$ over $R$ (or $R_j$). Hence the elements of $R[G]$ (or $R_j[G]$) are of the form $\sum_{i=1}^{n} a_{g_i}g_i$ where $a_{g_i} \in R$ (or $\sum_{i=1}^{n} a_{g_i}^{(j)}g_i$ where $a_{g_i}^{(j)} \in R_j$). It is clear that the map $\Lambda: R[G]\rightarrow R^n$, defined by $\sum_{i=1}^{n} a_{g_i}g_i\rightarrow (a_{g_1},a_{g_2},\ldots
,a_{g_n})$, is a $R$-modules isomorphism.

 We define  two operations over  $R_1[G]\times\ldots\times R_s[G]$:
$$(\sum_{i=1}^{n} a_{g_i}^{(1)}g_i,\ldots,\sum_{i=1}^{n} a_{g_i}^{(s)}g_i)+(\sum_{i=1}^{n} b_{g_i}^{(1)}g_i,\ldots,\sum_{i=1}^{n} b_{g_i}^{(s)}g_i)~~~~~~~~~~~~~~~~~~~~~~$$$$~~~~~~~~~~~~~~~~~~~~~~~~~=(\sum_{i=1}^{n} (a_{g_i}^{(1)}+b_{g_i}^{(1)})g_i,\ldots,\sum_{i=1}^{n} (a_{g_i}^{(s)}+b_{g_i}^{(s)})g_i).$$
and
$$(\sum_{i=1}^{n} a_{g_i}^{(1)}g_i,\ldots,\sum_{i=1}^{n} a_{g_i}^{(s)}g_i)\cdot(\sum_{i=1}^{n} b_{g_i}^{(1)}g_i,\ldots,\sum_{i=1}^{n} b_{g_i}^{(s)}g_i)~~~~~~~~~~~~~~~~~~~~~~$$$$~~~~~~~~~~~~~~~~~~~~~~~~~=(\sum_{i=1}^{n}(\sum _{j=1}^{n}a_{g_j}^{(1)}b_{g_j^{-1}g_i}^{(1)})g_i,\ldots,\sum_{i=1}^{n}(\sum _{j=1}^{n}a_{g_j}^{(s)}b_{g_j^{-1}g_i}^{(s)})g_i).$$
where $a_{g_i}^{(j)}, b_{g_i}^{(j)} \in R_j$ for all $1\leq j \leq s$.

It is easy to prove that  the $R_1[G]\times\ldots\times R_s[G]$  is an algebra.

\begin{thm}\label{theorem3.5}
Let $R= R_1\times R_2\times\cdots\times R_s$ is a principal ideal ring where $R_i$ is a chain ring for $1\leq i\leq s$. If $G$ is a finite group,
then
$$R[G]\cong R_1[G]\times\ldots\times R_s[G].$$
\end{thm}
\pf Suppose that $G=\{g_1,\ldots,g_n\}$. Then we define a map $\Phi$ from $R[G]$ to $R_1[G]\times\ldots\times R_s[G]$ as follows:
$$
\Phi:R[G]\longrightarrow R_1[G]\times\ldots\times R_s[G]$$
$$~~~~~~~~~~~~~~~\sum_{i=1}^{n} r_{g_i}g_i\longrightarrow(\sum_{i=1}^{n} r_{g_i}^{(1)}g_i,\ldots,\sum_{i=1}^{n} r_{g_i}^{(s)}g_i),$$
 where $r_{g_i}=(r_{g_i}^{(1)},\ldots,r_{g_i}^{(s)})\in R$ and $r_{g_i}^{(i)}\in R_i$ for $1\leq i\leq s$.

Clearly, $\Phi$ is a bijection.

For any $\sum_{i=1}^{n} a_{g_i}g_i$ and $\sum_{i=1}^{n} b_{g_i}g_i$ in $R[G]$, we have
$$\Phi(\sum_{i=1}^{n} a_{g_i}g_i+\sum_{i=1}^{n} b_{g_i}g_i)=\Phi(\sum_{i=1}^{n} (a_{g_i}+b_{g_i})g_i)~~~~~~~~~~~~~~~~~~~~~~~~~~~~~~~~~~~~$$
$$~~~~~~~~~~~~~~=(\sum_{i=1}^{n} (a_{g_i}^{(1)}+b_{g_i}^{(1)})g_i,\ldots,\sum_{i=1}^{n} (a_{g_i}^{(s)}+b_{g_i}^{(s)})g_i)$$
$$=\Phi(\sum_{i=1}^{n} a_{g_i}g_i)+\Phi(\sum_{i=1}^{n} b_{g_i}g_i).$$
and
$$\Phi(\sum_{i=1}^{n} a_{g_i}g_i\cdot\sum_{j=1}^{n} b_{g_j}g_j)=\Phi(\sum_{i=1}^{n}(\sum _{j=1}^{n}a_{g_j}b_{g_j^{-1}g_i})g_i)~~~~~~~~~~~~~~~~~~~~~~~~~~~~~~~~~~~~~~~~~~~$$
$$~~~~~~~~~~~~~~~~~~~~~~~=(\sum_{i=1}^{n}(\sum _{j=1}^{n}a_{g_j}b_{g_j^{-1}g_i})^{(1)}g_i,\ldots,\sum_{i=1}^{n} (\sum _{j=1}^{n}a_{g_j}b_{g_j^{-1}g_i})^{(s)}g_i)$$
$$~~~~~~~~~~~~~~=(\sum_{i=1}^{n}(\sum _{j=1}^{n}a_{g_j}^{(1)}b_{g_j^{-1}g_i}^{(1)})g_i,\ldots,\sum_{i=1}^{n}(\sum _{j=1}^{n}a_{g_j}^{(s)}b_{g_j^{-1}g_i}^{(s)})g_i)$$
$$=\Phi(\sum_{i=1}^{n} a_{g_i}g_i)\cdot\Phi(\sum_{i=1}^{n} b_{g_i}g_i).~~~~~~~~~~~~~$$
\qed

From now on, we denote the inverse of the map $\Phi$  by $\mathrm{CRT}$. The above theorem can be rewritten in the following form.

\begin{thm}\label{th:3.5}
Let $R=R_1\times R_2\times\cdots\times R_s$ is a principal ideal ring where $R_i$ is a chain ring for $1\leq i\leq s$. If $G$ is a finite group,
then
$$R[G]= \mathrm{CRT}(R_1[G],\ldots, R_s[G]).$$
\end{thm}

\section{LCP of group codes over finite principal ideal rings}
A right ideal of $R[G]$ (or $R_j[G]$) is called a group code
over $R$ (or $R_j$) (see \cite{Cem1} for group codes over finite chain rings). Throughout this section,  ideals will be $2$-sided
and they will be referred to as group codes.

Two group codes $C_1$ and $C_2$ over $R[G]$ (or $R_j[G]$) are permutation equivalent provided there is a permutation of
coordinates which sends $C_1$ to $C_2$. Then two group codes $C_1$ and $C_2$ are permutation equivalent if and only if there is a permutation matrix $P$ such that $C_2=C_1P$, where $C_1P= \{\mathbf{y} | \mathbf{y} = \mathbf{x}P~ \mathrm{for}~\mathbf{x} \in C_1\}$.

Let $C_j$ be a  group code over $R_j[G]$ for all $1\leq j \leq s$,  and let
$$C = \mathrm{CRT}(C_1,C_2,\ldots,C_s)= \Phi^{-1}(C_1 \times C_2\times\cdots\times C_s)$$
$$= \{\Phi^{-1}(\mathbf{c}_1, \mathbf{c}_2,\ldots, \mathbf{c}_s)|\mathbf{c}_j \in C_j\}.$$
We call $C$ the Chinese product of group codes $C_1,C_2,\ldots,C_s$.

\begin{thm}\label{th:3.6} Let $C_j$ be a  group code over the  $R_j[G]$ for all $1\leq j \leq s$. Then
$C=\mathrm{CRT}(C_1,C_2,\ldots,C_s)$ is a  group code over the  $R[G]$.
\end{thm}
\pf For any $\mathbf{a}\in C$, there is a $\mathbf{a}_j\in C_j$ such that
$$\mathbf{a}=\Phi^{-1}(\mathbf{a}_1,\ldots,\mathbf{a}_s),$$
where $\mathbf{a}_j=\sum_{i=1}^{n}a_{g_i}^{(j)}g_i$ with $a_{g_i}^{(j)}\in R_j$ for $1\leq j\leq s$.

Suppose that $a_{g_i}=(g_{g_i}^{(1)},g_{g_i}^{(2)},\ldots,g_{g_i}^{(s)})$ for $1\leq i\leq n$. Then $\mathbf{a}=\sum_{i=1}^{n}a_{g_i}g_i$.
Therefore, for any $g\in G$, we have  $g\mathbf{a}=\sum_{i=1}^{n}a_{g_i}gg_i=\sum_{i=1}^{n}a_{g^{-1}g_i}g_i$.

On the other hand, $g\mathbf{a}_j=\sum_{i=1}^{n}a_{g_i}^{(j)}gg_i=\sum_{i=1}^{n}a_{g^{-1}g_i}^{(j)}g_i$ for $1\leq j\leq s$.

Thus, we have $g\mathbf{a}=\Phi^{-1}(g\mathbf{a}_1,\ldots,g\mathbf{a}_s)$. Since $C_j$ is  a  group code over $R_j[G]$, $g\mathbf{a}_j\in C_j$. Thus, $g\mathbf{a}\in C$.

By using a similar technique we can show that $\mathbf{a}g\in C$.

Summarizing, we have proved that $C$ is an ideal in $R[G]$, i.e., $C$ is a  group code over  $R[G]$.
\qed

Now, we  give a useful lemma that will be used in later characterization of LCP  of group codes in $R[G]$.
\begin{lem}\label{le:3.4} Let $C_j$ and $D_j$ be two  group codes over the  $R_j[G]$ for all $1\leq j \leq s$. If
$C=\mathrm{CRT}(C_1,C_2,\ldots,C_s)$ and $D=\mathrm{CRT}(D_1,D_2,\ldots,D_s)$, then

$(1)$ $C\cap D=\mathrm{CRT}(C_1\cap D_1,C_2\cap D_2,\ldots,C_s\cap D_s)$;

$(2)$ $C+D=\mathrm{CRT}(C_1+ D_1,C_2+D_2,\ldots,C_s+ D_s)$.
\end{lem}
\pf  Suppose that $\mathbf{a}=\sum_{i=1}^{n}a_{g_i}g_i$ where $a_{g_i}=(a_{g_i}^{(1)},\ldots,a_{g_i}^{(s)})$ and $a_{g_i}^{(j)}\in R_j$ for $1\leq j\leq s$. Then $\mathbf{a}\in R[G]$.

$(1)$ $ \mathbf{a}\in C\cap D$ if and only if $$\mathbf{a}=\Phi^{-1}(\sum_{i=1}^{n}a_{g_i}^{(1)}g_i,\sum_{i=1}^{n}a_{g_i}^{(2)}g_i,\ldots,\sum_{i=1}^{n}a_{g_i}^{(s)}g_i)\in\mathrm{CRT}(C_1,C_2,\ldots,C_s),$$
and  $$\mathbf{a}=\Phi^{-1}(\sum_{i=1}^{n}a_{g_i}^{(1)}g_i,\sum_{i=1}^{n}a_{g_i}^{(2)}g_i,\ldots,\sum_{i=1}^{n}a_{g_i}^{(s)}g_i)\in\mathrm{CRT}(D_1,D_2,\ldots,D_s).$$
Then, $ \mathbf{a}\in C\cap D$ if  and only if $\mathbf{a}\in\mathrm{CRT}(C_1\cap D_1,C_2\cap D_2,\ldots,C_s\cap D_s)$.

Therefore
$$C\cap D=\mathrm{CRT}(C_1\cap D_1,C_2\cap D_2,\ldots,C_s\cap D_s).$$

$(2)$ $ \mathbf{a}\in C + D$ if and only if $$\mathbf{a}=\Phi^{-1}(\sum_{i=1}^{n}a_{g_i}^{(1)}g_i,\sum_{i=1}^{n}a_{g_i}^{(2)}g_i,\ldots,\sum_{i=1}^{n}a_{g_i}^{(s)}g_i)+\Phi^{-1}(\sum_{i=1}^{n}b_{g_i}^{(1)}g_i,\sum_{i=1}^{n}b_{g_i}^{(2)}g_i,\ldots,\sum_{i=1}^{n}b_{g_i}^{(s)}g_i)$$
$$=\Phi^{-1}(\sum_{i=1}^{n}a_{g_i}^{(1)}g_i+\sum_{i=1}^{n}b_{g_i}^{(1)}g_i,\sum_{i=1}^{n}a_{g_i}^{(2)}g_i+\sum_{i=1}^{n}b_{g_i}^{(2)}g_i,\ldots,\sum_{i=1}^{n}a_{g_i}^{(s)}g_i+\sum_{i=1}^{n}b_{g_i}^{(s)}g_i),$$
where $$\Phi^{-1}(\sum_{i=1}^{n}a_{g_i}^{(1)}g_i,\sum_{i=1}^{n}a_{g_i}^{(2)}g_i,\ldots,\sum_{i=1}^{n}a_{g_i}^{(s)}g_i)\in C,$$
and
$$\Phi^{-1}(\sum_{i=1}^{n}b_{g_i}^{(1)}g_i,\sum_{i=1}^{n}b_{g_i}^{(2)}g_i,\ldots,\sum_{i=1}^{n}b_{g_i}^{(s)}g_i)\in D.$$
Then, $ \mathbf{a}\in C+ D$ if  and only if $\mathbf{a}\in\mathrm{CRT}(C_1+ D_1,C_2+D_2,\ldots,C_s+ D_s)$.

Therefore
$$C+ D=\mathrm{CRT}(C_1+ D_1,C_2+ D_2,\ldots,C_s+ D_s).$$
\qed

The proof of the following lemma is similar the proofs of Theorems $2.4$, $2,7$ and Lemma $2.5$ in \cite{Doug}, so we omit it here.
\begin{lem}\label{le:3.5}. Let $C=\mathrm{CRT}(C_1,C_2,\ldots,C_s)$ be a group code over $R[G]$, where $C_j$  is a  group code over the  $R_j[G]$ for $1\leq j\leq s$. Then

$(1)$ $|C|=\prod_{j=1}^s|C_j|$;

$(2)$ $C^{\perp}=\mathrm{CRT}(C_1^{\perp},C_2^{\perp},\ldots,C_s^{\perp})$.
\end{lem}

The following result gives a necessary and sufficient condition for a pair $(C, D)$ of group codes over group algebra $R[G]$ to be LCP.

\begin{thm}\label{th:3.7} Let $C_j$ and $D_j $ be group codes in $R_j[G]$ for all $1\leq j\leq s$, and let
$C=\mathrm{CRT}(C_1,C_2,\ldots,C_s)$ and $D=\mathrm{CRT}(D_1,D_2,\ldots,D_s)$.  Then $(C, D)$ is LCP of group codes in $R[G]$ if and only if $(C_j, D_j)$ is LCP of group codes in $R_j[G]$ for all $1\leq j\leq s$.
\end{thm}
\pf  Since $(C_j, D_j)$ is LCP of group codes in $R_j[G]$ for all $1\leq j\leq s$, we have $C_j\cap D_j=\{0\}$ and $C_j+ D_j=R_j[G]$ or $|C_j||D_j| =|R_j[G]|$ for all $1\leq j\leq s$. By Lemma \ref{le:3.4},
$$C\cap D=\mathrm{CRT}(C_1\cap D_1,C_2\cap D_2,\ldots,C_s\cap D_s)=\mathrm{CRT}(\mathbf{0},\mathbf{0},\ldots,\mathbf{0})=\{\mathbf{0}\}.$$
Then, according Lemma \ref{le:3.5} $(1)$,
$$|C+D|=|C|\cdot|D|=\prod_{j=1}^s|C_j|\cdot\prod_{j=1}^s|D_j|=\prod_{j=1}^s|C_j||D_j|=\prod_{j=1}^s|R_j[G]|=|R[G]|.$$
Therefore, $(C, D)$ is LCP of group codes in $R[G]$.

Conversely, suppose that $(C, D)$ is  LCP of group codes in $R[G]$.  Then $C+D=R[G]$ and $C\cap D=\{0\}$. By Theorem \ref{th:3.5} and Lemma \ref{le:3.4}, we have
$$\mathrm{CRT}(C_1\cap D_1,C_2\cap D_2,\ldots,C_s\cap D_s)=\{\mathbf{0}\},$$
and
$$\mathrm{CRT}(C_1+ D_1,C_2+ D_2,\ldots,C_s+ D_s)=\mathrm{CRT}(R_1[G],R_2[G],\ldots,R_s[G]).$$
Thus,
$$C_1\cap D_1=\{\mathbf{0}\},C_2\cap D_2=\{\mathbf{0}\},\ldots,C_s\cap D_s=\{\mathbf{0}\},$$
$$C_1+ D_1=R_1[G],C_2+ D_2=R_2[G],\ldots,C_s+ D_s=R_s[G].$$
This proves that $(C_j, D_j)$ is LCP of group codes in $R_j[G]$ for all $1\leq j\leq s$.
\qed

In order to prove that $C$ and $D^{\perp}$ are equivalent codes, we give the following lemma which can be found in \cite{Cem1}.

\begin{lem}\label{le:3.8}Let $(\widetilde{C}, \widetilde{D})$ be an LCP of group codes in $\widetilde{R}[\widetilde{G}]$, where $\widetilde{R}$ is a finite chain ring
and $\widetilde{G}$ is a finite group. Then $\widetilde{C}$ and $\widetilde{D}^{\perp}$ are equivalent codes.
\end{lem}

\begin{thm}\label{th:3.8}  Let $R=\mathrm{CRT}(R_1,R_2,\ldots,R_s)$ be a finite principal ideal ring, where $R_j$ is a finite chain ring for all $1\leq j\leq s$,  and let $G$ be a finite group. If $(C, D)$ is an LCP of group codes in $R[G]$, Then $C$ and $D^{\perp}$ are equivalent codes. In
particular, $d(D^{\perp})=d(C)$.
\end{thm}
\pf Let $C=\mathrm{CRT}(C_1,C_2,\ldots,C_s)$ and $D=\mathrm{CRT}(D_1,D_2,\ldots,D_s)$, where $C_j$ and $D_j$ are  group codes in $R_j[G]$ for all $1\leq j\leq s$. Since $(C, D)$ is an LCP of group codes in $R[G]$,  $(C_j,D_j)$ is an  LCP of group codes in $R_j[G]$ for all $1\leq j\leq s$ by Theorem \ref{th:3.7} $(1)$.

According Lemma \ref{le:3.8}, $C_j$ and $D_j^{\perp}$ are equivalent codes for all $1\leq j\leq s$. Then there is a permutation matrix $P_j$ such  that $C_j=D_j^{\perp}P_j$ for all $1\leq j\leq s$.

Set $$P=\begin{pmatrix}P_1&0&\cdots&0\\0&P_2&\cdots&0\\\vdots&\vdots&\cdots&\vdots\\0&0&\cdots&P_s\end{pmatrix}.$$
Then $C_1\times C_2 \times \cdots\times C_s=(D_1^{\perp}\times D_2^{\perp} \times \cdots\times D_s^{\perp})P$.

Since $C=\mathrm{CRT}(C_1,C_2,\ldots,C_s)\cong C_1\times C_2 \times \cdots\times C_s$ and $D^{\perp}=\mathrm{CRT}(D_1^{\perp},D_2^{\perp},\ldots,D_s^{\perp})$, we have
$$C=\mathrm{CRT}(C_1,C_2,\ldots,C_s)=P ~\mathrm{CRT}(D_1^{\perp},D_2^{\perp},\ldots,D_s^{\perp}).$$
Thus, $C$ and $D^{\perp}$ are equivalent codes.
\qed

\section*{Acknowledgements}
This work was supported by Research Funds of Hubei Province, Grant
No. D20144401.


\begin{thebibliography}{99}

\bibitem{Bor} Borello M., de Cruz J., Willems W.: A note on linear complementary pairs of group codes. Discret. Math. $\mathbf{343}$, 111905 (2020).
\bibitem{Carlet} Carlet C.,  G$\ddot{u}$neri C.,  Mesnager S.,  $\ddot{O}$zbudak F.: Construction of some codes suitable for both side channel and fault injection attacks,  Proceedings of International Workshop on the Arithmetic of Finite Fields (WAIFI 2018), Bergen(2018).
\bibitem {Carlet1}Carlet C.,  G$\ddot{u}$neri C., $\ddot{O}$zbudak F., $\ddot{O}$zkaya B., Sol$\grave{e}$ P.: On linear complementary pairs of codes,  IEEE Trans. Inform. Theory, $\mathbf{64}$(1),6583-6588(2018).
\bibitem {Cem} G$\ddot{u}$neri C., $\ddot{O}$zkaya B., Say${\i}$c${\i}$ S.: On linear complementary pair of $nD$ cyclic codes. IEEE Commun. Lett.
$\mathbf{22}$, 2404-2406 (2018).
\bibitem {Cem1} G$\ddot{u}$neri C.,  Mart${\i}$nez-Moro E., Say${\i}$c${\i}$  S.: Linear complementary pair of group codes over finite chain rings, Designs, Codes and Cryptogr. https://doi.org/10.1007/s10623-020-00792-1
\bibitem{Doug} Dougherty S. T.,  Liu H.: Independence of vectors in codes over rings. Des. Codes Cryptogr. {\bf51}, 55-68(2009).
\bibitem{Doug1}  Dougherty S. T., Kim, J.L.,  Kulosman H.:  MDS codes over finite principal ideal rings. Des. Codes Cryptogr. $\mathbf{50}$, 77-92(2009)
\bibitem{Ling}Ling S., Sol$\acute{e}$ P.: On the algebraic structure of quasi-cyclic codes II: Chain rings. Des. Codes Cryptogr. $\mathbf{30}$, 113-130(2003).
\bibitem{Liu} Liu X. S., Liu H.: LCD codes over finite chain rings, Finite Field Appl. $\mathbf{15}$, 1-19(2015).
\bibitem{Ngo} Ngo X. T., Bhasin S., Danger J.-L., Guilley S., Najm Z.: Linear complementary dual code improvement to strengthen encoded circuit against hardware Trojan horses, in Proc. IEEE Int. Symp. Hardw. Oriented Secur. Trust (HOST), 82-87( 2015).
\bibitem {Nor}Norton G. H.,  S$\check{a}$l$\check{a}$gean A. S.: On the structure of linear and cyclic codes over a finite chain ring. Applicable
algebra in engineering, communication and computing, $\mathbf{10}$, 489-506(2000).
\end{thebibliography}
\end{document}